\documentclass[11pt]{article}
\usepackage{axodraw}
\usepackage{amssymb,latexsym}
\usepackage{graphicx}
\usepackage{amssymb, amsmath, amsopn, amsthm,slashed}
\usepackage[all]{xy}
\renewcommand{\theequation}{\thesection.\arabic{equation}}
\newcounter{saveeqn}
\newcommand{\add}{\addtocounter{equation}{1}}
\newcommand{\alpheqn}{\setcounter{saveeqn}{\value{equation}}%
\setcounter{equation}{0}%
\renewcommand{\theequation}{\mbox{\thesection.\arabic{saveeqn}{\alph{equation}}}}}
\newcommand{\reseteqn}{\setcounter{equation}{\value{saveeqn}}%
\renewcommand{\theequation}{\thesection.\arabic{equation}}}

\begin{document}
\begin{titlepage}
\thispagestyle{empty}
\begin{flushright}
    SU-ITP-2007-02\\
    SLAC-PUB-12367\\
\today
\end{flushright}
\vspace{30pt}
\begin{center}
    { \LARGE{\textbf{Constraints on Meta-stable de Sitter}\\ \vspace{0.7cm} \textbf{Flux Vacua}}}

    \vspace{40pt}

  {\large
   Masoud Soroush\ \footnote{\ soroush@stanford.edu}}

    \vspace{10pt}

    \vspace{10pt} { \textsc{Department of Physics and SLAC\\
    Stanford University\\ Stanford, CA 94305-94309, USA}}
    \vspace{20pt}
 \end{center}
\begin{abstract}
\vspace{0.5cm}
We consider flux compactification of type IIB string theory as the orientifold limit of an F-theory on a Calabi-Yau fourfold. We show that when supersymmetry is dominantly broken by the axion-dilaton and the contributions of the F-terms associated with complex structure moduli are small, the Hessian of the flux potential always has tachyonic modes for de Sitter vacua. This implies that there exist no meta-stable de Sitter vacua in this limit. Moreover, we find that the stability requirement imposes a relation between the values of cosmological constant and the scale of supersymmetry breaking for non-supersymmetric anti de Sitter vacua in this limit. The proof is general and does not rely on the details of the geometry of the compact Calabi-Yau internal space. We finally analyze the consequences of these constraints on the statistics of meta-stable de Sitter vacua and address some other related issues.
\end{abstract}
\vspace{10pt}
\end{titlepage}
\tableofcontents
\section{Introduction}
In order to establish concrete cosmological models which explain inflation, dark energy, and capture the other realistic features that cosmological data require, we need to provide a detailed account of de Sitter vacua. String theory which naturally accommodates gravity beside other known forces of nature produces a vast class of de Sitter vacua via compactification down to four dimensions. In a generic compactification (at large volume and weak coupling), one is allowed to turn on various fluxes in the compact dimensions. The existence of fluxes strongly affects the physics of the four dimensions (for a comprehensive review on the subject, see \cite{Douglas:2006es} and \cite{Douglas:2003um}).
\par
In type IIB string theory compactified on a Calabi-Yau orientifold, the allowed fluxes are those associated with RR and NS three-form field strengths. The resulting low energy effective theory is ${\mathcal{N}}=1$ four dimensional supergravity whose tree level action has the no scale structure \cite{Giddings:2001yu}. The fluxes stabilize the axion-dilaton field and complex structure moduli of the Calabi-Yau but not the K\"{a}hler moduli. According to KKLT scenario, in order to complete the process and fix all moduli, one needs to incorporate the non-perturbative effects to the superpotential \cite{Kachru:2003aw}. In this manner, one can start with the axion-dilaton and complex structure moduli and stabilize them by fluxes at the first step and then introduce the non-perturbative effects to the superpotential and stabilize the K\"{a}hler moduli, as well as the other moduli. Having all moduli stabilized, there are two possible ways to construct de Sitter vacua. In the first approach, one starts with a meta-stable supersymmetric (the minimization equations for supersymmetric vacua are easier to solve) anti de Sitter vacuum and at the end introduces a stack of anti D3-branes to lift the vacuum to a de Sitter one \cite{Kachru:2003aw}. Of course, this way requires some amount of fine-tuning in order not to destabilize the vacuum. Taking this approach, it is crucial to know about the statistics of meta-stable supersymmetric vacua. It has been shown in \cite{Ashok:2003gk} that the statistics of these vacua is nicely governed by the K\"{a}hler and Ricci forms of the moduli space\footnote{\ The moduli space of the theory in this case is the Cartesian product of complex structure moduli space of Calabi-Yau and the moduli space of elliptic curves, which is a K\"{a}hler manifold.} and this has been supported in \cite{Giryavets:2004zr}, \cite{Denef:2004ze}, and \cite{Denef:2004cf},  by considering explicit examples.
\par
The second approach to construct a de Sitter vacuum has been proposed and explored for an explicit example in \cite{Saltman:2004sn}. In this case, one starts with a meta-stable de Sitter vacuum with stabilized axion-dilaton and complex structure moduli by turning on the fluxes. In the next step, non-perturbative effects are included and the K\"{a}hler moduli get stabilized. The benefit of this method is that it does not need an uplifting process, because we have simply started by a de Sitter minimum. Of course, this strategy is not particular to type IIB string theory compactification and has also been applied to heterotic M-theory vacua \cite{Curio:2006dc}. In this case, the complex structure moduli are first stabilized at a high scale by $H$-flux and, in the next step, the K\"{a}hler moduli are stabilized by incorporating non-perturbative effects coming from different sources, namely the gaugino condensation, $M2$ and $M5$ instantons.
\par
Taking this approach, the crucial thing is then studying and analyzing the statistics of meta-stable de Sitter vacua before the K\"{a}hler moduli are stabilized. Various aspects of this subject have been investigated in \cite{Denef:2004ze} and \cite{Denef:2004cf}. The results shows that the construction of meta-stable de Sitter vacua (even before involving the stabilization of K\"{a}hler moduli) is hard. We briefly present  the following observations from \cite{Denef:2004ze} and \cite{Denef:2004cf}. The simplest example is first to consider only one complex structure modulus. In the following two limits, the stability of the flux vacua has been studied.
\begin{itemize}
\item \textbf{Large Complex Structure Limit}: In the limit of large complex structure, the only existing Yukawa coupling of special geometry is constant and furthermore its covariant derivative vanishes. Therefore, the explicit computation which is needed to check the stability of the non-supersymmetric vacua (in particular de Sitter vacua) is considerably simplified. The Hessian of the potential is a $4\times4$ matrix and thus we can easily  diagonalize it in terms of its parameters. The result in \cite{Denef:2004ze} shows that for de Sitter vacua, at least, one of the eigenvalues of the Hessian of the potential is always negative in all branches of solutions. This proves that the Hessian of the potential is not positive definite and hence it is impossible to obtain any meta-stable de Sitter vacuum in the limit of large complex structure.
\item \textbf{Conifold Limit}: One might naively think that the result which was obtained in the previous limit (large complex structure) is not a general property of de Sitter vacua and if one considers other examples, then it might be possible to find meta-stable de Sitter vacua. In \cite{Denef:2004ze}, another example, which is more complicated but still calculable, has been investigated. In this example, a Calabi-Yau with only one complex structure is considered near a generic conifold degeneration. In the vicinity of the origin of the moduli space which is parameterized by the period of the vanishing cycle of the Calabi-Yau, we only deal with the singular terms and drop all analytic terms. Unlike the previous example, the Yukawa coupling of special geometry is not a constant term in this case and further it has a non-vanishing covariant derivative. If we then proceed and calculate the Hessian of the potential at the locus of critical points and find all meta-stable non-supersymmetric vacua in this limit, as it has been done in \cite{Denef:2004ze}, we will find that all those non-supersymmetric vacua are anti de Sitter. Therefore, once again, we find that there are no meta-stable de Sitter vacua\footnote{Nevertheless, it is discussed in \cite{Hebecker:2006bn} that if one allows fine-tuning of fluxes, the tachyonic modes of the mass matrix disappear for two or more number of complex structure moduli in the conifold limit.}.
\end{itemize}
It should be emphasized that the two mentioned limits are not two single examples and, in fact, each of them forms a family of examples. Because, we did not specified the Calabi-Yau entirely and we only required that it should possess only one complex structure modulus. There are many known Calabi-Yau threefolds which satisfy this condition \cite{Hosono:1993qy}.
\par
So far, we may think that this somewhat strange property of de Sitter vacua is due to the fact that we have only  considered one complex structure modulus. Perhaps, it will be possible to find meta-stable de Sitter vacua if we allow  Calabi-Yau threefolds which have more complex structure moduli. Unfortunately, the full analysis of the situation is not possible because of various kinds of computational complications which arise by adding more complex structure moduli. Therefore, in order to investigate this new situation, we have to restrict ourselves to a non-generic simple example. In \cite{Denef:2004cf}, this situation for an orientifold example has been investigated.
\begin{itemize}
\item \textbf{Case of the Orientifold $T^{6}/{\mathbb{Z}}_{2}^{2}$}\ : This orbifold has three complex structure moduli, one for each torus $T^{2}$. The only independent non-vanishing component of special geometry Yukawa coupling is the one whose all three indices are different and is constant. Moreover, the covariant derivative of this Yukawa coupling vanishes. Therefore, this example is relatively simple and can be analyzed exactly. The result in \cite{Denef:2004cf} shows that there are no meta-stable de Sitter vacua (and even no approximately Minkowski vacua) and the mass matrix always has a negative eigenvalue. Of course, the situation can be modified by adding an uplifting D-term, as the authors propose in \cite{Denef:2004cf}. But before involving any uplifting procedure, it is impossible to find meta-stable de Sitter vacua.
\end{itemize}
The above observations suggest that perhaps there is a deeper fact which prohibits the existence of meta-stable de Sitter vacua (at least for some regions of moduli space) and we have missed it. If such a fact exists, it should probably be independent of the characteristics and details of the compact Calabi-Yau internal space.
\par
Of course, the difficulty of constructing meta-stable de Sitter vacua is not particular to flux compactification models and it has also been analyzed in the context of supergravity. In \cite{Gomez-Reino:2006dk}, the existance of meta-stable non-supersymmetric vacua has been studied for ${\mathcal{N}}=1$ supergravity models and it has been shown that the stability requirement leads to certain constrains on the curvature of the moduli spcae of these theories and the implications of these constraints have been explored for specific examples in \cite{Gomez-Reino:2006wv}. More interestingly, we exactly know why it is difficult to construct those vacua in ${\mathcal{N}}=2$ supergravity models. In the absence of hypermultiplets in ${\mathcal{N}}=2$ supergravity, it was first found in \cite{deWit:1984pk} that the holomorphic-antiholomorphic part of the mass matrix is always tachyonic (it has, at least, one negative eigenvalue) for de Sitter vacua. This implies that the mass matrix is not positive definite and therefore, any de Sitter vacuum is perturbatively unstable. The proof of this theorem, which has clearly been explained in \cite{Fre:2002pd}, only relies on the special geometry structure of the vector multiplet. Consequently, if we consider four-dimensional ${\mathcal{N}}=2$ supergravity as an embedding theory of type IIB string theory compactified on a Calabi-Yau threefold (without turning the fluxes on), then this theorem states that it is impossible to construct any meta-stable de Sitter vacuum regardless of the details of the topology and geometry of the Calabi-Yau manifold. The main purpose of this paper is to show that, in fact, similar theorems exist for flux compactifications in certain limits and, at least, partially answer the earlier question of why constructing stable de Sitter vacua is very difficult. In this manner, we exclude some regions of moduli space for which there is no hope to find  meta-stable de Sitter vacua.
\par
This paper is organized as follows. In section 2, we introduce the essential ingredients and backgrounds for F-theory flux compactification on a Calabi-Yau fourfold. We then take the Denef-Douglas model of landscape of string vacua, in which one neglects the K\"{a}hler moduli and restricts oneself to complex structures and the axion-dilaton. In fact, we ignore the no scale structure of supergravity \cite{Giddings:2001yu}. Nevertheless, the analysis of the landscape of vacua of this model is important if we consider the supersymmetric vacua of this model as the starting point of KKLT construction \cite{Kachru:2003aw} and de Sitter vacua of this model for the starting point of the scenario which was proposed in \cite{Saltman:2004sn}. In either scenarios \cite{Kachru:2003aw}, \cite{Saltman:2004sn}, the no scale structure of supergravity is ultimately broken and the K\"{a}hler moduli get stabilized as well as complex structures. If the construction of de Sitter vacua is possible, then the proposal in \cite{Saltman:2004sn} may even seem more attractive than the KKLT construction, since it does not require adding any anti D3-branes at the end to uplift the anti de Sitter vacuum to a de Sitter one.
\par
In section 3, we find the Hessian of the flux potential. We organize different sectors of the Hessian of the potential such that it becomes a Hermitian matrix. In section 4, we discuss the stability of the critical points of the potential in the limit where supersymmetry is dominantly broken by the contribution of the axion-dilaton. The reason we have to restrict ourselves to a limit is that if we consider the simplest example of this type in which there exists only one complex structure modulus, we find that even this example is complicated enough that no result can be extracted in general, without considering any limit. Therefore, we need to analyze the problem in different regions of moduli space separately. In this paper, we focus on this specific limit (the dilaton dominated limit) because of two reasons. First, to obtain a de Sitter vacuum without D-term uplifting, we should allow supersymmetry breaking parameters ($DW$) to compete with the value of the superpotential. Secondly, it has been proved in \cite{Brignole:1993dj} that all soft supersymmetry breaking terms are universal in the dilaton dominated limit and do not depend on any specific feature of the model.
\par
In section 4.1, for stable non-degenerate critical points, we first require the positivity of the Hessian of the potential at the critical points. Since the Hessian of the potential is a Hermitian matrix, it is necessary for any principal submatrix of the Hessian to be positive definite. Therefore, as a necessary but not sufficient condition the holomorphic-antiholomorphic sector of the Hessian must be positive definite by itself. However, we find one the eigenvalues of this sector of the Hessian is proportional to the minus of a mixture of cosmological constant and supersymmetry breaking scale. This implies that it would be impossible to find meta-stable de Sitter vacua in the  regions of the moduli space which correspond to dilaton dominated limit. Moreover, the stability requirement gives us a bound for the scale of supersymmetry breaking for non-supersymmetric meta-stable anti de Sitter vacua in this limit. Using the similarity between the mathematical structure of flux vacua and non-supersymmetric black hole attractors, we argue, in section 4.2, that degenerate critical points cannot be stable for generic fluxes, although it might be possible to achieve some meta-stable de Sitter vacua by fine-tuning of parameters. We will investigate the consequences of the obtained results on the statistics of meta-stable de Sitter vacua in section 5. We show that the limit we considered in section 4 defines certain regions of moduli space which should eventually be excluded when looking for meta-stable de Sitter vacua. In section 6, we explain that this stability condition  is true only for non-supersymmetric vacua and is not applicable for supersymmetric vacua. Therefore, this point might be useful for the current attempts for finding appropriate criteria to distinguish between the landscape of string vacua and the swampland \cite{Ooguri:2006in}. Section 7 is devoted to discussions and concluding thoughts.
\section{Backgrounds of Flux Compactification}
\setcounter{equation}{0}
In this section, we briefly review F-theory flux compactification on an elliptically fibered Calabi-Yau fourfold $X$ in the orientifold limit in which $X=(Y\times T^{2})/{\mathbb{Z}}_{2}$, where $Y$ is a Calabi-Yau threefold. In the language of type IIB string theory, this is equivalent to compactifying on the orientifold limit of $Y$ with a constant axion-dilaton. The resulting low energy four-dimensional effective theory is ${\mathcal{N}}=1$ supergravity whose information is encoded by a flux superpotential $W$ and a K\"{a}hler potential $K$. Throughout the paper, we assume that both flux superpotential $W$ and K\"{a}hler potential $K$ only depend on complex structure moduli of $X$. We denote the complex structure moduli space of $X$ as ${\mathcal{M}}$ which is a complex K\"{a}hler manifold. Because of the product structure of $X$, ${\mathcal{M}}$ is identified with ${\mathcal{M}}={\mathcal{M}}_{cs}(Y)\times{\mathcal{E}}$, where ${\mathcal{M}}_{cs}(Y)$ is the moduli space of complex structures of $Y$ and ${\mathcal{E}}$ is the moduli space of elliptic curves which corresponds to the space of possible values for the axion-dilaton $\tau$. Another way of saying this is that we have ignored the K\"{a}hler moduli of $Y$. The reason we can do this is that if we incorporate the non-perturbative effects which are present in string theory \cite{Kachru:2003aw}, the no-scale structure of the tree level action of supergravity is then broken and K\"{a}hler moduli can essentially be stabilized.
\par
The potential term in effective Lagrangian of ${\mathcal{N}}=1$ supergravity theory, in the Planckian units set equal to one, is given by
\begin{eqnarray}\label{pot}
V=e^{K}\Big(\sum_{A=0}^{h^{2,1}(Y)}|D_{A}W|^{2}-3|W|^{2}\Big)\ ,
\end{eqnarray}
where $A=0$ refers to axion-dilaton $\tau$ and $A=a\in\{1,\cdots,h^{2,1}(Y)\}$ to complex structures of $Y$. The covariant derivative is constructed with respect to the K\"{a}hler connection of a line bundle ${\mathcal{L}}$ (whose first Chern class is $[c_{1}({\mathcal{L}})]=[\omega]$, where $\omega$ is a K\"{a}hler form defined on ${\mathcal{M}}$) over ${\mathcal{M}}$ and they are contracted via the Weil-Petersson metric derived from the K\"{a}hler potential.
\\
The K\"{a}hler potential of the four-dimensional theory is given by
\begin{eqnarray}\label{kahler-pot}
K=-\ln(i\langle\Omega_{4},\bar{\Omega}_{4}\rangle)=-\ln(i\langle\Omega_{1},\bar{\Omega}_{1}\rangle)
-\ln(i\langle\Omega_{3},\bar{\Omega}_{3}\rangle)\ ,
\end{eqnarray}
where $\Omega_{4}$ is a nowhere vanishing holomorphic four form defined on $X$. In the orientifold limit, we can easily express $\Omega_{4}$ in terms of an appropriate holomorphic three form $\Omega_{3}$ of the Calabi-Yau threefold $Y$ and the holomorphic one form $\Omega_{1}$ of the torus $T^{2}$ as $\Omega_{4}=\Omega_{1}\wedge\Omega_{3}$. Doing so, the K\"{a}hler potential takes the form of the r.h.s. of (\ref{kahler-pot}). To be more explicit, we can choose a basis for the space of harmonic one forms on $T^{2}$ and rewrite $\Omega_{1}$ in terms of that basis. Suppose $\alpha,\beta\in{\mathcal{H}}^{1}(T^{2})$ are canonical harmonic one forms on $T^{2}$ and form a basis for ${\mathcal{H}}^{1}(T^{2})$ such that we can rewrite $\Omega_{1}=\beta-\tau\alpha$. Then it is easy to see that
\begin{eqnarray}\label{holo1form}
D_{0}\Omega_{1}={\mathcal{C}}_{0}\bar{\Omega}_{1}\ ,\ \ \ D_{0}D_{0}\Omega_{1}
=(\partial_{0}+\partial_{0}K_{1}-\Gamma^{0}_{00})(\partial_{0}+\partial_{0}K_{1})\Omega_{1}=0\ ,
\end{eqnarray}
where ${\mathcal{C}}_{0}=i\langle\Omega_{1},\partial_{0}\Omega_{1}\rangle=-\frac{1}{\tau-\bar{\tau}}$, $K_{1}$ is that part of K\"{a}hler potential which comes from the torus $K_{1}=-\ln(i\langle\Omega_{1},\bar{\Omega}_{1}\rangle)$, and finally $\Gamma^{0}_{00}=G^{0\bar{0}}\partial_{\bar{0}}G_{\bar{0}0}$ is the Levi-Civita connection for the K\"{a}hler metric on the torus. The complex structure moduli space of $Y$, ${\mathcal{M}}_{cs}(Y)$, is a K\"{a}hler manifold which enjoys special geometry \cite{Strominger:1990pd}. If we define a holomorphic symmetric three-tensor ${\mathcal{C}}\in\Gamma({\mathcal{L}},(T{\mathcal{M}}_{cs}(Y))^{\otimes3})$ componentwise as ${\mathcal{C}}_{abc}=i\langle\Omega,\partial_{a}\partial_{b}\partial_{c}\Omega\rangle$, then the special geometry structure of ${\mathcal{M}}_{cs}(Y)$ tells us
\begin{eqnarray}\label{holo3form}
D_{a}D_{b}\Omega_{3}={\mathcal{C}}_{abc}\bar{D}^{c}\bar{\Omega}_{3}\ .
\end{eqnarray}
In the physics literature, ${\mathcal{C}}_{abc}$ is often known as the special geometry Yukawa coupling. Furthermore, special geometry gives a relation between the curvature of the tangent bundle of ${\mathcal{M}}_{cs}(Y)$ and the Yukawa couplings as
\begin{eqnarray}\label{yakawacurv}
{\mathcal{R}}_{a\bar{b}c\bar{d}}=G_{a\bar{b}}G_{c\bar{d}}+G_{a\bar{d}}G_{c\bar{b}}-e^{2K_{3}}
G^{e\bar{f}}{\mathcal{C}}_{ace}\bar{\mathcal{C}}_{\bar{b}\bar{d}\bar{f}}\ ,
\end{eqnarray}
where $K_{3}=-\ln(i\langle\Omega_{3},\bar{\Omega}_{3}\rangle)$ is the K\"{a}hler potential on ${\mathcal{M}}_{cs}(Y)$ and $G_{a\bar{b}}$ is the corresponding K\"{a}hler metric.
\par
The flux superpotential $W$ is defined as a section of the line bundle ${\mathcal{L}}$ by \cite{Gukov:1999ya}, \cite{Douglas:2005df}
\begin{eqnarray}\label{super-pot}
W=\langle G_{4},\Omega_{4}\rangle\equiv \int_{X} G_{4}\wedge\Omega_{4}\ ,
\end{eqnarray}
where $G_{4}\in H^{4}(X,{\mathbb{Z}})$ is the quantized four form flux and can be written as $G_{4}=-\alpha\wedge F_{3}+\beta\wedge H_{3}$, in which the harmonic three forms $F_{3}$ and $H_{3}$ are identified with RR and NS-NS type IIB fluxes respectively.
\par
Given a complex structure for the fourfold $X$, we have a natural Hodge decomposition as
\begin{eqnarray}\label{hodge-dec}
H^{4}(X,{\mathbb{C}})=H^{4,0}(X,{\mathbb{C}})\oplus H^{3,1}(X,{\mathbb{C}})\oplus H^{2,2}(X,{\mathbb{C}})\oplus
H^{1,3}(X,{\mathbb{C}})\oplus H^{0,4}(X,{\mathbb{C}})\ .
\end{eqnarray}
Since $X$ has a product structure, we can determine the dimension of each of the above subspaces in terms of the Hodge numbers of the Calabi-Yau threefold $Y$ and the torus $T^{2}$. Using the Kunneth decomposition formula\footnote{\ If $M_{1}$ and $M_{2}$ are two complex manifolds, then the following isomorphism is held
\begin{eqnarray}\label{Kunneth1}
H^{u,v}(M_{1}\times M_{2})\cong\bigoplus_{\tiny{\begin{array}{c}
                                            p+r=u \\
                                            q+s=v
                                          \end{array}}
}H^{p,q}(M_{1})\otimes H^{r,s}(M_{2})\ .
\end{eqnarray}
As an immediate consequence for dimensions of the cohomology spaces, we have the following relation between the Hodge numbers of $M_{1}$, $M_{2}$, and $M_{1}\times M_{2}$
\begin{eqnarray}\label{Kunneth2}
h^{u,v}(M_{1}\times M_{2})=\sum_{\tiny{
\begin{array}{c}
p+r=u \\
q+s=v
\end{array}}} h^{p,q}(M_{1})h^{r,s}(M_{2})\ .
\end{eqnarray}}, we easily find $h^{4,0}(X)=h^{0,4}(X)=1$, $h^{3,1}(X)=h^{1,3}(X)=1+n$, and $h^{2,2}(X)=2n$, where $n=h^{2,1}(Y)$ is the dimension of the complex structure moduli space of $Y$. We can go further and find an explicit basis for each of these subspaces. Due to Griffiths transversality (more details can be found in the appendix), covariant derivatives of any $(p,q)$-form $\Psi$ defined on $X$ ($\Psi\in\Omega^{p,q}(X)$) belong to
\begin{eqnarray}\label{griffith}
D_{A_{1}}\cdots D_{A_{k}}\Psi\in \bigoplus_{j=0}^{k} H^{p-j,q+j}(X,{\mathbb{C}})\ .
\end{eqnarray}
Therefore, the covariant derivative of the holomorphic four form $\Omega_{4}$ has two pieces $D_{A}\Omega_{4}=\chi^{4,0}+\chi^{3,1}$, where $\chi^{4,0}=C\Omega_{4}$. However, the $(4,0)$ part vanishes because  $C=\langle\bar{D}_{\bar{A}}\bar{\Omega}_{4},\Omega_{4}\rangle
=-\langle\bar{\Omega}_{4},\bar{D}_{\bar{A}}\Omega_{4}\rangle
=-\langle\bar{\Omega}_{4},\bar{\partial}_{\bar{A}}\Omega_{4}\rangle=0$ and hence, $D_{A}\Omega_{4}$ is a purely $(3,1)$ form. Using (\ref{griffith}), we can express $D_{A}D_{B}\Omega_{4}=\zeta^{3,1}+\zeta^{2,2}$. With the same technique, we can see that $\zeta^{3,1}$ vanishes since $\langle\bar{D}_{\bar{C}}\bar{\Omega}_{4},D_{A}D_{B}\Omega_{4}\rangle=D_{A}\langle\bar{D}_{\bar{C}}\bar{\Omega}_{4}
,D_{B}\Omega_{4}\rangle-\langle D_{A}\bar{D}_{\bar{C}}\bar{\Omega}_{4},D_{B}\Omega_{4}\rangle=-D_{A}G_{B\bar{C}}-
\langle G_{A\bar{C}}\bar{\Omega}_{4},D_{B}\Omega_{4}\rangle=0$. This then implies that $D_{A}D_{B}\Omega_{4}$ is a purely $(2,2)$ form. We can use covariant derivatives of $\Omega_{4}$ to form basis for cohomology spaces (\ref{hodge-dec}) as following
\begin{eqnarray}\label{basiscoh}
&&H^{4,0}(X,{\mathbb{C}})=\mbox{span}\{\Omega_{4}\}\ ,\ H^{0,4}(X,{\mathbb{C}})=\mbox{span}\{\bar{\Omega}_{4}\}\ ,\nonumber\\
&&H^{3,1}(X,{\mathbb{C}})=\mbox{span}\{D_{A}\Omega_{4};\ A=0\cdots n\}\ ,\ H^{1,3}(X,{\mathbb{C}})=\mbox{span}\{\bar{D}_{\bar{A}}
\bar{\Omega}_{4};\ A=0\cdots n\}\ ,\nonumber\\
&&H^{2,2}(X,{\mathbb{C}})=\mbox{span}\ \{D_{0}D_{a}\Omega_{4};\ a=1\cdots n\}\cup\{\bar{D}_{\bar{0}}\bar{D}_{\bar{a}}\bar{\Omega}_{4};\ a=1\cdots n\}\ .
\end{eqnarray}
In this case, it is quite easy to see that the flux form $G_{4}$ has the following Hodge decomposition
\begin{eqnarray}\label{G-dec}
G_{4}&=&\bar{W}\Omega_{4}-(\bar{D}^{A}\bar{W})D_{A}\Omega_{4}+(\bar{D}^{0}\bar{D}^{a}\bar{W})D_{0}D_{a}\Omega_{4}
\nonumber\\
&&+(D_{0}D_{a}W)\bar{D}^{0}\bar{D}^{a}\bar{\Omega}_{4}-(D_{A}W)\bar{D}^{A}\bar{\Omega}_{4}+W\bar{\Omega}_{4}\ .
\end{eqnarray}
Furthermore, the flux form $G_{4}$ is subject to the tadpole cancelation condition
\begin{eqnarray}\label{tadpole}
L\equiv\langle G_{4}\wedge G_{4}\rangle=\frac{\chi(X)}{24}-(N_{D3}-N_{\overline{D}3})\ ,
\end{eqnarray}
where $N_{D3}$ and $N_{\overline{D}3}$ are the number of D3 and anti-D3 branes. Of course, $L$ cannot be made indefinitely large by adding any number of anti-D3 branes since the system will decay into a system of flux and D3 branes system.
\section{Hessian of the Flux Potential}
\setcounter{equation}{0}
In order to check the stability of the critical points of the flux potential (\ref{pot}), we first need to find the Hessian of the potential. In this section, we find all elements of the Hessian. Since we want to argue about the stability of the vacua at the position of the critical points after all, we calculate the second order covariant derivatives of the potential instead of the ordinary derivatives.
\par
It is clear that because of (\ref{holo1form}) and (\ref{holo3form}), all second order covariant derivatives of the superpotential are not independent of each other and we have the following relations
\begin{eqnarray}
D_{0}D_{0}W&=&D_{0}D_{0}\langle G_{4},\Omega_{4}\rangle=\langle G_{4},(D_{0}D_{0}\Omega_{1})\wedge\Omega\rangle=0\ ,\label{DDW1}\\
D_{a}D_{b}W&=&D_{a}D_{b}\langle G_{4},\Omega_{4}\rangle=\langle G_{4},\Omega_{1}\wedge D_{a}D_{b}\Omega\rangle=
{\mathcal{C}}_{abc}\bar{D}^{c}\langle G_{4},\Omega_{1}\wedge\bar{\Omega}\rangle\nonumber\\
&=&{\mathcal{C}}_{0}{\mathcal{C}}_{abc}\bar{D}^{0}\bar{D}^{c}\langle G_{4},\bar{\Omega}_{4}\rangle=
{\mathcal{F}}_{abc}\bar{D}^{0}\bar{D}^{c}\bar{W}\ ,\label{DDW2}
\end{eqnarray}
where we have defined ${\mathcal{F}}_{abc}\equiv{\mathcal{C}}_{0}{\mathcal{C}}_{abc}$. Another useful formula is the commutation relation between holomorphic and antiholomorphic covariant derivatives. We can prove a generic formula for this case and use it whenever it is needed to commute covariant derivatives acting on a general object. Assuming  $P\in\Omega^{m,n}({\mathcal{M}})$ which has the holomorphic-antiholomorphic K\"{a}hler weights $(h_{P},\bar{h}_{P})$, the commutation relation of $D_{a}$ and $\bar{D}_{\bar{b}}$ acting on $P$ is then given by
\begin{eqnarray}\label{DDbar}
[\bar{D}_{\bar{b}},D_{a}]P_{c_{1}\cdots c_{m}\bar{d}_{1}\cdots\bar{d}_{n}}&=&(h_{P}-\bar{h}_{P})g_{a\bar{b}}
P_{c_{1}\cdots c_{m}\bar{d}_{1}\cdots\bar{d}_{n}}-\sum_{i=1}^{m}{\mathcal{R}}^{e_{i}}_{a\bar{b}c_{i}}P_{c_{1}\cdots e_{i}\cdots c_{m}\bar{d}_{1}\cdots\bar{d}_{m}}\nonumber\\
&&-\sum_{j=1}^{n}{\mathcal{R}}^{\bar{e}_{j}}_{\bar{b}a\bar{d}_{j}}
P_{c_{1}\cdots c_{m}\bar{d}_{1}\cdots\bar{e}_{j}\cdots\bar{d}_{n}}\ ,
\end{eqnarray}
where ${\mathcal{R}}$ is the curvature of the tangent bundle of the moduli space ${\mathcal{M}}$ as the base manifold. We notice that if we exchange the position of $D_{a}$ and $\bar{D}_{\bar{b}}$ in the above relation, then the positions  of $h_{P}$ and $\bar{h}_{P}$ are also exchanged in the r.h.s. of (\ref{DDbar}), so that the skew-symmetry property of the commutator is preserved. Putting all these relations together, we are now able to compute all derivatives of the potential. We first calculate the first order derivatives of the flux potential (\ref{pot})
\begin{eqnarray}
&&\partial_{0}V=D_{0}V=e^{K}\Big((D_{0}D_{a}W)\bar{D}^{a}\bar{W}-2(D_{0}W)\bar{W}\Big)\ ,\label{dV1}\\
&&\partial_{a}V=D_{a}V=e^{K}\Big({\mathcal{F}}_{abc}(\bar{D}^{b}\bar{W})\bar{D}^{0}\bar{D}^{c}\bar{W}
+(\bar{D}^{0}\bar{W})D_{0}D_{a}W-2\bar{W}D_{a}W\Big)\label{dV2}\ ,
\end{eqnarray}
where we have used (\ref{DDW1}) in deriving (\ref{dV1}), and (\ref{DDW2}) together with a special case of (\ref{DDbar}) acting on $W$ with K\"{a}hler weight (1,0) for (\ref{dV2}). Now, we want to compute second order derivatives of the potential which form the elements of the Hessian. Since we are interested to evaluate the Hessian of the potential at the position of a critical point and because $D_{A}D_{B}V(p_{0})=\partial_{A}\partial_{B}V(p_{0})$ holds for any critical point $p_{0}\in{\mathcal{M}}$ ($D_{A}V(p_{0})=0$), we can instead calculate the second order covariant derivatives of the potential, which manifestly respect the covariant form of the equations. The Hessian of the potential is a square $2(h^{2,1}(Y)+1)$ dimensional Hermitian matrix
\begin{eqnarray}\label{hessian}
{\mathcal{H}}_{V}(p_{0})=\left(
                           \begin{array}{cc}
                             D_{A}\bar{D}^{B}V(p_{0}) & D_{A}D_{B}V(p_{0}) \\
                             \bar{D}^{A}\bar{D}^{B}V(p_{0}) & \bar{D}^{A}D_{B}V(p_{0}) \\
                           \end{array}
                         \right)\ .
\end{eqnarray}
Using (\ref{dV1}), (\ref{dV2}), and other techniques we introduced in this section, we can calculate all elements of the Hessian of the potential. These relations have also been computed in \cite{Denef:2004ze} in an orthonormal frame (tangent space) of the moduli space, which admits a diagonal flat metric. The holomorphic-holomorphic elements of the Hessian are given by
\begin{eqnarray}
&& D_{0}D_{0}V=0\ ,\label{ddV1}\\
&& D_{0}D_{a}V=D_{a}D_{0}V=e^{K}\Big({\mathcal{F}}_{abc}(\bar{D}^{b}\bar{W})\bar{D}^{c}\bar{W}
-\bar{W}D_{0}D_{a}W\Big)\ ,\label{ddV2}\\
&& D_{a}D_{b}V=e^{K}\Big((D_{a}{\mathcal{F}}_{bcd})(\bar{D}^{c}\bar{W})\bar{D}^{0}\bar{D}^{d}\bar{W}
+2{\mathcal{F}}_{abc}(\bar{D}^{0}\bar{W})\bar{D}^{c}\bar{W}-{\mathcal{F}}_{abc}\bar{W}\bar{D}^{0}\bar{D}^{c}\bar{W}
\Big)\ .\label{ddV3}
\end{eqnarray}
We do not need these elements of the Hessian for the rest of the paper and they have been mentioned only for completeness purposes. However, we do need the holomorphic-antiholomorphic elements of the Hessian for our purpose and they are given by
\begin{eqnarray}
&& D_{0}\bar{D}^{0}V=e^{K}\Big(-2|W|^{2}-2|D_{0}W|^{2}+|D_{a}W|^{2}+|D_{0}D_{a}W|^{2}\Big)\ ,\label{ddV4}\\
&& D_{0}\bar{D}^{a}V=e^{K}\Big(-(D_{0}W)\bar{D}^{a}\bar{W}+\bar{\mathcal{F}}^{abc}(D_{0}D_{b}W)(D_{0}D_{c}W)\Big)\ ,\label{ddV5}\\
&& D_{a}\bar{D}^{0}V=e^{K}\Big(-(D_{a}W)\bar{D}^{0}\bar{W}+{\mathcal{F}}_{abc}(\bar{D}^{0}\bar{D}^{b}\bar{W})
(\bar{D}^{0}\bar{D}^{c}\bar{W})\Big)\ ,\label{ddV6}\\
&& D_{a}\bar{D}^{b}V=e^{K}\Big(-(2|W|^{2}+|D_{0}W|^{2})\delta^{b}_{a}-2(D_{a}W)\bar{D}^{b}\bar{W}
+(D_{0}D_{a}W)\bar{D}^{0}\bar{D}^{b}\bar{W}\nonumber\\
&& \hspace{2.6cm} +{\mathcal{F}}_{acd}\bar{\mathcal{F}}^{dbe}\big((\bar{D}^{c}\bar{W})D_{e}W
+(\bar{D}^{0}\bar{D}^{c}\bar{W})D_{0}D_{e}W\big)\Big)\ .\label{ddV7}
\end{eqnarray}
We notice that the above expressions for the elements of the Hessian, (\ref{ddV1})-(\ref{ddV7}), are generic and they are valid everywhere (not only at critical points).
\section{Stability of the Critical Points of the Flux Potential}
\setcounter{equation}{0}
After finding the critical points of the flux potential, we should find out whether they are perturbatively stable. In other words, we should make sure that any vacuum is a local minimum in each direction of the moduli space. If the Hessian of the potential is tachyonic and has at least one negative eigenvalue at a critical point, then the critical point is not a local minimum and therefore, it cannot be regarded as a meta-stable vacuum. We consider degenerate and non-degenerate critical points of the potential separately in the following two subsections.
\par
Here, in section 4.1, we assume that the contribution to supersymmetry breaking parameter is dominated by the axion-dilaton and the contribution of F-terms associated with complex structure moduli is small in comparison with it. More precisely, what we assume is that $|D_{0}W|\gg|D_{a}W|$. Later in section 5, we discuss the consequences of this limit and will figure out that it corresponds to particular regions in moduli space. In fact, we will show that it is impossible to find any meta-stable de Sitter vacuum in those regions of moduli space which correspond to this limit and therefore they should be  excluded.
\par
It is interesting to notice that in the context of four dimensional heterotic string theory, this condition is well motivated by reasons coming from phenomenology. In \cite{Brignole:1993dj}, a string theory motivated parametrization of supersymmetry breaking parameters of the supersymmetric standard model is provided. In a simple example, it is assumed that there is only one F-term in addition to the one associated with the axion-dilaton. The goldstino angle is then defined as the ratio of $\tan\theta=\frac{|D_{0}W|}{|D_{1}W|}$. The limit $\sin\theta\rightarrow1$ corresponds to dilaton dominated supersymmetry breaking. The important feature of dilaton dominated limit is that all soft parameter terms are universal and independent of the details of the model. As $\sin\theta$ decreases, the model dependence of the soft terms increases and the resulting soft terms may not be universal anymore. This construction is, of course, in the framework of heterotic string. For the case of type IIB string theory, one can consult \cite{Camara:2003ku} for details of the dilaton domination.
\subsection{Non-degenerate Critical Points}
In this section, we work with non-degenerate critical points of the potential. In the next part, we will discuss the stability of degenerate critical points. According to (\ref{dV1}) and (\ref{dV2}), for any critical point of the potential ($D_{A}V=0$) we have
\begin{eqnarray}
&&(D_{0}D_{a}W)\bar{D}^{a}\bar{W}=2\bar{W}D_{0}W\ ,\label{dV=01}\\
&&{\mathcal{F}}_{abc}(\bar{D}^{b}\bar{W})\bar{D}^{0}\bar{D}^{c}\bar{W}=2\bar{W}D_{a}W-(\bar{D}^{0}\bar{W})D_{0}D_{a}W\ .\label{dV=02}
\end{eqnarray}
We notice that any quantity in this section is evaluated at the position of a non-degenerate critical point $p_{0}\in{\mathcal{M}}$ of the potential $V$, for which we have ($D_{A}V(p_{0})=0$ and $\mbox{det}({\mathcal{H}}_{V}(p_{0}))\neq0$). We drop the letter $p_{0}$ to show that the value at the critical point but it should automatically be understood.
\par
If a non-degenerate critical point of the potential is a local minimum (meta-stable), then the Hessian of the potential must be a positive definite matrix at the position of the critical point (the Morse index of the critical point should be zero.). Since ${\mathcal{H}}_{V}$ is a Hermitian matrix, it is positive definite if and only if every principal submatrix of ${\mathcal{H}}_{V}$ is positive definite\footnote{\ In fact, if $H$ is a finite dimensional Hermitian matrix, then the following statements are equivalent:
\begin{itemize}
\item $H$ is positive definite.
\item Every principal submatrix of $H$ is positive definite.
\item Every principal sub-determinant of $H$ is positive.
\item Every leading principal sub-determinant of $H$ is positive.
\end{itemize}}. In particular, if we consider the holomorphic-antiholomorphic block of the Hessian, which is a square $(h^{2,1}(Y)+1)$ dimensional leading principal submatrix of ${\mathcal{H}}_{V}(p_{0})$, then this block by itself must be positive definite. Now, we want to investigate whether the holomorphic-antiholomorphic part of the Hessian is positive definite at the position of the critical point $p_{0}$.
\par
In the limit where $|D_{0}W|\gg|D_{a}W|$, the elements of the holomorphic-antiholomorphic part of the Hessian of the potential, (\ref{ddV4})-(\ref{ddV7}), are given by
\begin{eqnarray}
&& D_{0}\bar{D}^{0}V=e^{K}\Big(-2|W|^{2}-2|D_{0}W|^{2}+|D_{0}D_{a}W|^{2}\Big)\ ,\label{ddV-limit1}\\
&& D_{0}\bar{D}^{a}V=e^{K}\Big(-(D_{0}W)\bar{D}^{a}\bar{W}+\bar{\mathcal{F}}^{abc}(D_{0}D_{b}W)(D_{0}D_{c}W)\Big)\ ,\label{ddV-limit2}\\
&& D_{a}\bar{D}^{0}V=e^{K}\Big(-(D_{a}W)\bar{D}^{0}\bar{W}+{\mathcal{F}}_{abc}(\bar{D}^{0}\bar{D}^{b}\bar{W})
(\bar{D}^{0}\bar{D}^{c}\bar{W})\Big)\ ,\label{ddV-limit3}\\
&& D_{a}\bar{D}^{b}V=e^{K}\Big(-(2|W|^{2}+|D_{0}W|^{2})\delta^{b}_{a}+(D_{0}D_{a}W)\bar{D}^{0}\bar{D}^{b}\bar{W}\nonumber\\
&&\hspace{2.6cm}+{\mathcal{F}}_{acd}\bar{\mathcal{F}}^{dbe}(\bar{D}^{0}\bar{D}^{c}\bar{W})D_{0}D_{e}W\Big)\ .\label{ddV-limit4}
\end{eqnarray}
In above expressions, we have kept the first order terms in $D_{a}W$, but we have neglected second order terms ${\mathcal{O}}(|D_{a}W|^{2})$. Now, we claim that $D_{A}W$ is an eigenvector of the the holomorphic-antiholomorphic part of the Hessian of the potential at the critical point. By proving this statement, i. e. $(D_{A}\bar{D}^{B}V)D_{B}W=\lambda D_{A}W$, we will also determine the corresponding eigenvalue $\lambda$. First, we start with the zeroth component of the eigenvalue equation
\begin{eqnarray}\label{eigen1}
(D_{0}\bar{D}^{B}V)D_{B}W&=&(D_{0}\bar{D}^{0}V)D_{0}W+(D_{0}\bar{D}^{b}V)D_{b}W\nonumber\\
&=&e^{K}\Big(-2(|W|^{2}+|D_{0}W|^{2})D_{0}W+2W(D_{0}D_{b}W)\bar{D}^{b}\bar{W}\Big)\nonumber\\
&=&2e^{K}\Big(|W|^{2}-|D_{0}W|^{2}\Big)D_{0}W\ .
\end{eqnarray}
In first line of the above equation (\ref{eigen1}), we have substituted the expressions (\ref{ddV-limit1}) and (\ref{ddV-limit2}) for holomorphic-antiholomorphic elements of the Hessian. We have also used (\ref{dV=02}) in the second line and (\ref{dV=01}) in order to get the last line. In the second line, we also get a term $|D_{a}W|^{2}$ which is negligible in this limit. For the $a$-th component of the eigenvalue equation, we have
\begin{eqnarray}\label{eigen2}
(D_{a}\bar{D}^{B}V)D_{B}W&=&(D_{a}\bar{D}^{0}V)D_{0}W+(D_{a}\bar{D}^{b}V)D_{b}W\nonumber\\
&=&e^{K}\Big(-2(|W|^{2}+|D_{0}W|^{2})D_{a}W+{\mathcal{F}}_{abc}(D_{0}W)(\bar{D}^{0}\bar{D}^{b}\bar{W})
\bar{D}^{0}\bar{D}^{c}\bar{W}\nonumber\\
&&+(D_{0}D_{a}W)(\bar{D}^{0}\bar{D}^{b}\bar{W})D_{b}W+{\mathcal{F}}_{acd}(\bar{D}^{0}\bar{D}^{c}\bar{W})
(\bar{\mathcal{F}}^{dbe}(D_{0}D_{e}W)D_{b}W)\Big)\nonumber\\
&=&e^{K}\Big(-2(|W|^{2}+|D_{0}W|^{2})D_{a}W+2W(\bar{D}^{0}\bar{W})D_{0}D_{a}W\nonumber\\
&&+2W{\mathcal{F}}_{acd}(\bar{D}^{0}\bar{D}^{c}\bar{W})\bar{D}^{d}\bar{W}\Big)\nonumber\\
&=&2e^{K}\Big(|W|^{2}-|D_{0}W|^{2}\Big)D_{a}W\ .
\end{eqnarray}
We have substituted (\ref{ddV-limit3}) and (\ref{ddV-limit4}) for the r.h.s. of the second equality. If we use (\ref{dV=01}) and (\ref{dV=02}) for the the last two terms of the second equality respectively, we will obtain r.h.s. of the third equality which results the last line, using (\ref{dV=02}) again. Therefore, we find that $D_{A}W$ is, indeed, one of the eigenvectors of the holomorphic-antiholomorphic part of the Hessian at the position of the critical points of the potential with corresponding eigenvalue $\lambda=2e^{K}(|W|^{2}-|D_{0}W|^{2})$. If we rewrite the eigenvalue $\lambda$ in terms of the critical value of potential $V=e^{K}(|D_{0}W|^{2}-3|W|^{2})$, we have
\begin{eqnarray}\label{eigen3}
(D_{A}\bar{D}^{B}V)D_{B}W=-2(V+2e^{K}|W|^{2})D_{A}W=-2(\Lambda+2M_{3/2}^{2})D_{A}W\ ,
\end{eqnarray}
where $\Lambda$ is the cosmological constant (the critical value of $V$) and $M_{3/2}=e^{K/2}|W|$ is the mass of gravitino (scale of supersymmetry breaking). First, it should be noticed that this equation does not give us any information about supersymmetric vacua, because the eigenvector $D_{A}W$ will be the zero vector for supersymmetric vacua and thus (\ref{eigen3}) is trivially satisfied. However, this relation (eq. (\ref{eigen3})) has important consequences for non-supersymmetric vacua. It implies that
\begin{itemize}
\item for any de Sitter vacuum ($\Lambda>0$), the holomorphic-antiholomorphic part of the Hessian of the potential has a negative definite eigenvalue and therefore, the complete Hessian is not a positive definite matrix. This means that any de Sitter vacuum is perturbatively unstable.
\item any non-supersymmetric Minkowski vacuum ($\Lambda=0$ and $W\neq0$) is also unstable. We notice that this statement is not valid when the superpotential vanishes at locus of the critical point. Because in this case the vacuum is supersymmetric and for supersymmetric vacua, (\ref{eigen3}) is trivially satisfied.
\item for any non-supersymmetric meta-stable anti de Sitter ($\Lambda<0$), there is relation between the value of the cosmological constant and the scale of supersymmetry breaking\footnote{\ This relation is valid before involving any uplifting procedure. As soon as we lift anti de Sitter vacua (by any possible method), this relation is destroyed.}. In fact, there is a lower bound for the value of the supersymmetry breaking scale in any meta-stable anti de Sitter vacuum: $2M_{3/2}^{2}<|\Lambda|$. The highest possible value for $M_{3/2}$ is determined by the value of cosmological constant in that anti de Sitter vacuum, which is, of course, in agreement with the physical intuition. Furthermore, it should be noticed that this bound is not sufficient in order to get meta-stable anti de Sitter vacua, but it is absolutely necessary. Another interesting feature of this bound is that demanding small value of cosmological constant is correlated with having low scale of supersymmetry breaking.
\end{itemize}
\subsection{Degenerate Critical Points}
In this section, we argue that degenerate critical points are not perturbatively stable for a generic set of fluxes. For degenerate critical points, the determinant of Hessian of the potential (\ref{hessian}) vanishes at the locus of the critical points and we have to take higher order perturbations into account. More precisely, we have to consider cubic terms as the leading correction along the degenerate directions of the moduli space. If cubic terms are non-vanishing, then the critical point is neither a local minimum nor a local maximum but rather an inflection point. However, if all cubic terms also vanish along the degenerate directions of moduli space, we have a chance to find a local minimum (or maximum) by  quartic terms of the expansion.
\par
The same phenomenon happens in the context of non-supersymmetric black hole attractors. The mathematical structure of  these black holes, which appear as four-dimensional solutions of ${\mathcal{N}}=2$ supergravity as the low energy effective theory of type IIA string theory compactified on a Calabi-Yau threefold, is very similar to the mathematical structure governed on flux compactification \cite{Kallosh:2005ax}. It turns out that supersymmetric black holes are perturbatively stable as a result of special geometry. However, this statement is no longer true for non-supersymmetric black holes and one needs to check the stability of the critical points explicitly. It has been shown in \cite{Kallosh:2006ib} non-supersymmetric black hole solutions for generic electric and magnetic charges which correspond to charges of even branes wrapping on appropriate cycles are degenerate critical points of the black hole potential. Therefore, to check the stability of these solutions, it is necessary to consider the higher order terms. It is shown \cite{Kallosh:2006ib} that the cubic terms do not vanish for a generic set of electric and magnetic charges and thus the solutions are not stable in general. But if we consider a specific set of charges, for example if only D2-D6 brane charges exist (or D0-D4 brane charges which is, in fact, dual to D2-D6 brane system by electric-magnetic duality), then all cubic terms in the expansion of the black hole potential vanish. Hence for this specific configuration (D2-D6 brane system), one should consider the quartic terms in the expansion. The result shows \cite{Kallosh:2006ib} that all coefficients of quartic terms in the expansion along degenerate directions of the moduli space are positive and eventually non-supersymmetric solutions of D2-D6 brane system are stable.
\par
By analogy between the mathematical structure of flux compactification and black hole attractors, we conclude that for generic set of fluxes there is no reason to believe that the cubic terms should vanish in the expansion of the flux potential in terms of moduli fields around a degenerate critical point. This would imply that the degenerate critical point is an inflection point in some directions of moduli space and therefore it cannot be stable. Nevertheless, it might be possible to find stable degenerate critical points by choosing specific fluxes. In that situation, it is necessary for all coefficients of cubic terms to vanish along degenerate directions. Furthermore, all quartic terms must be positive in the expansion, as in the example of non-supersymmetric black hole attractors.
\section{Statistics of Meta-stable de Sitter Vacua}
\setcounter{equation}{0}
In this section, we analyze the consequences of the result we obtained in the previous section, namely (\ref{eigen3}), on the statistics of non-supersymmetric and, in particular, de Sitter vacua. But before doing that it would be instructive first to consider the analysis which has been done in \cite {Denef:2004cf} for the statistics of non-supersymmetric vacua in the regime of a small supersymmetry breaking parameter. As we will see in below, the stability requirement leads to a mild constraint in this regime whereas in our case it leads to a stringent constraint.
\par
It is shown in \cite{Denef:2004cf} that the critical points of the flux potential correspond to the eigenvectors of a square $2(n+1)$-dimensional Hermitian matrix $M$
\begin{eqnarray}\label{M-matrix}
M=\left(
    \begin{array}{cc}
      0 & e^{-i\theta}D_{A}D_{B}W \\
      e^{i\theta}\bar{D}_{\bar{A}}\bar{D}_{\bar{B}}\bar{W} & 0 \\
    \end{array}
  \right)\ ,
\end{eqnarray}
with the eigenvalue $2|W|$, in which $\theta$ is defined as the phase of the superpotential $e^{i\theta}=\frac{W}{|W|}$. Then the Hessian of the potential can be expanded in powers of $D_{A}W$ as
\begin{eqnarray}\label{d2VDW}
{\mathcal{H}}_{V}=(M+|W|\cdot{\mathbf{1}}_{2(n+1)})(M-2|W|\cdot{\mathbf{1}}_{2(n+1)})+{\mathcal{O}}(D_{A}W)+
{\mathcal{O}}((D_{A}W)^{2})\ ,
\end{eqnarray}
where the first term in above is the zeroth order term in $D_{A}W$ and the second and third terms are the first and second order terms of the expansion respectively. Now, we restrict ourselves to the region of moduli space ${\mathcal{M}}$ in which $D_{A}W$ is very small. This region of moduli space corresponds to almost supersymmetric solutions (See Figure 1). Without fine-tuning, these non-supersymmetric solutions are anti de Sitter vacua, because $W$ is of order of one (in Planckian units) for a generic flux, which implies $|W|\gg|D_{A}W|$. In other words, since in this limit, the value of the superpotential is dominant in the flux potential, it is not possible to obtain critical points with positive critical value of the potential and this has nothing to do with the stability requirement. As a result, there exists no de Sitter vacuum for generic fluxes in this limit unless one uplifts the vacua by introducing D-terms, as it has been proposed in \cite{Denef:2004cf}. Here, let us continue the analysis of these anti-de Sitter vacua without involving D-terms for uplifting.
\par
For this regime (almost supersymmetric vacua), we can ignore the first and second order terms in $D_{A}W$ in (\ref{d2VDW}). Therefore, the eigenvalues of $M$ can be regarded as {\textit{approximate}} eigenvalues of the Hessian. In this case, in order to analyze the stability of the vacua in this limit, it is sufficient to analyze the eigenvalues of $M$. Because of special form of $M$, (\ref{M-matrix}), the eigenvalues of $M$ appear as pairs of $(\lambda_{A},-\lambda_{A})$, where $\lambda_{A}\in{\mathbb{R}}^{+}$. We can reorder these eigenvalues such that $\lambda_{n}\geqslant\lambda_{n-1}\geqslant\cdots\geqslant\lambda_{0}\geqslant0$. According to (\ref{d2VDW}), the requirement to obtain a positive definite Hessian is that the lowest eigenvalue of $M$ should be greater than $2|W|$, namely $\lambda_{0}>2|W|$. Therefore, we only need to be worried about one eigenvalue. If $\lambda_{0}>2|W|$, then ${\mathcal{H}}_{V}$ will not have any tachyonic mode and the vacuum is perturbatively stable. In case $\lambda_{0}=2|W|$, the zeroth order term in (\ref{d2VDW}) vanishes and the first order term will determine whether or not the vacua are stable. The analysis in \cite{Denef:2004cf} shows that it requires some amount of fine-tuning.
\par
After all, since the stability of the vacua depends on the value of only one of the eigenvalues of matrix $M$, the probability to find a meta-stable non-supersymmetric vacuum is $\frac{1}{n+1}$. Notice that this estimate is much bigger than the naive probability one may think. Since $M$ has $2(n+1)$ eigenvalues, there is , at least for a uniform distribution of vacua, a fifty-fifty percent chance in order to get a positive eigenvalue in each direction of moduli space based on general grounds. Therefore, the probability to find a typical non-supersymmetric stable vacuum is $\frac{1}{2^{2(n+1)}}$ which is much less than $\frac{1}{n+1}$, as we already mentioned. Finally, the bottom line is that the stability requirement leads to a mild constraint on the number of non-supersymmetric vacua in the limit of very low supersymmetry breaking parameter.
\begin{figure}[!t]
\begin{center}
\begin{picture}(300,130)(-60,0)
\CCirc(50,30){5}{Gray}{Gray}
\CCirc(80,70){5}{Gray}{Gray}
\CCirc(100,10){5}{Gray}{Gray}
\CCirc(35,80){5}{Gray}{Gray}
\CCirc(150,50){5}{Gray}{Gray}
\CCirc(190,75){5}{Gray}{Gray}
\CCirc(185,20){5}{Gray}{Gray}
\CCirc(275,60){5}{Gray}{Gray}
\CCirc(50,30){1}{Black}{Black}
\CCirc(80,70){1}{Black}{Black}
\CCirc(100,10){1}{Black}{Black}
\CCirc(35,80){1}{Black}{Black}
\CCirc(150,50){1}{Black}{Black}
\CCirc(190,75){1}{Black}{Black}
\CCirc(185,20){1}{Black}{Black}
\CCirc(275,60){1}{Black}{Black}
\Oval(150,50)(65,160)(0)
\CArc(25,105)(60,240,300)
\CArc(25,-2)(57,66,114)
\CArc(128,150)(70,240,300)
\CArc(128,30)(60,66,114)
\CArc(250,80)(60,236,304)
\CArc(250,-29)(60,66,114)
\LongArrow(-135,-10)(-25,-10)
\LongArrow(-80,-10)(-80,100)
\CArc(-80,-10)(30,60,120)
\Line(-95,16)(-95,95)
\Line(-65,16)(-65,95)
\Line(-40,55)(-30,45)
\Line(-40,45)(-30,55)
\CCirc(-82,40){5}{Gray}{Gray}
\CCirc(-82,40){1}{Black}{Black}
\CCirc(-73,70){5}{Gray}{Gray}
\CCirc(-73,70){1}{Black}{Black}
\end{picture}
\end{center}
\vspace{0.8cm}
\caption{\textit{This figure shows the moduli space} ${\mathcal{M}}$ \textit{schematically. The left piece corresponds to the moduli space of elliptic curves}, ${\mathcal{E}}$, \textit{and the right piece represents the complex structure moduli space of the Calabi-Yau threefold}, ${\mathcal{M}}_{cs}(Y)$. \textit{The black points correspond to supersymmetric critical points for which} $D_{A}W=0$. \textit{The gray} $2$-\textit{dimensional disks in} ${\mathcal{E}}$ and $2n$-\textit{dimensional small balls in} ${\mathcal{M}}_{cs}(Y)$ \textit{represent regions of moduli space in the vicinity of  supersymmetric critical points of the flux potential.}}
\end{figure}
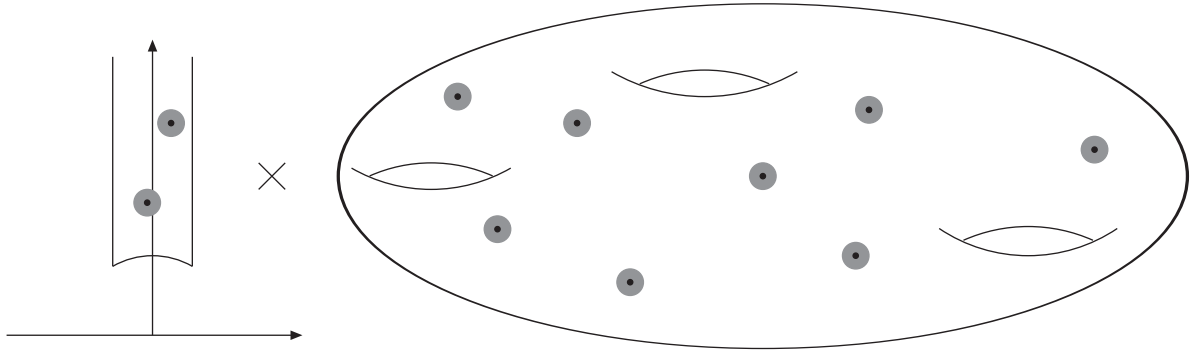
\par
Now, we are ready to analyze the statistics of meta-stable non-supersymmetric vacua in the limit where the supersymmetry breaking is dominated by axion-dilaton and the contribution of F-terms of complex structure moduli is small. Note that we are mostly interested in de Sitter vacua. In the previous limit, non-supersymmetric vacua are all anti de Sitter if we do not include D-term uplifting, but in this limit we can, in principle, have de Sitter vacua without incorporating D-terms. The first thing to notice is that clearly the procedure we took to analyze the statistics of vacua in the previous limit is spoiled for this new case, because the contribution of the supersymmetry breaking parameter is not negligible in this limit and the eigenvalues of the zeroth order term in (\ref{d2VDW}) cannot be considered as approximate eigenvalues of the Hessian anymore.
\par
First, let us to explore what regions of moduli space correspond to the limit of supersymmetry breaking dominated by axion-dilaton and small complex structure F-terms contribution. If we only allow small complex structure F-terms, then this implies that in the complex structure moduli space of the Calabi-Yau threefold, we have restricted ourselves to the vicinity of the position of supersymmetric critical points in ${\mathcal{M}}_{cs}(Y)$. But in the space of possible values of the axion-dilaton, ${\mathcal{E}}$ (remember that the whole moduli space ${\mathcal{M}}$ of the theory is ${\mathcal{M}}={\mathcal{E}}\times{\mathcal{M}}_{cs}(Y)$), we are far away from the position of supersymmetric critical points. For the almost supersymmetric solutions (the previous limit) in which we were restricted to the neighborhood of the supersymmetric critical points both in ${\mathcal{E}}$ and ${\mathcal{M}}_{cs}(Y)$, we found that there exists no meta-stable de Sitter vacuum for generic fluxes. Now, we can ask what happens if we still restrict ourself to small deformations around the position of supersymmetric critical points in ${\mathcal{M}}_{cs}(Y)$ but let the axion-dilaton freely vary in ${\mathcal{E}}$. This is the situation which has been represented in Figure 2. \par
In section 4.1, we showed that for non-degenerate critical points of the flux potential, the Hessian always has at least one tachyonic mode for de Sitter vacua. Therefore, non-degenerate critical points of the flux potential cannot be perturbatively stable de Sitter vacua in the limit where supersymmetry is dominantly broken by the axion-dilaton. For degenerate critical points, we argued in section 4.2 that they cannot be stable critical points for generic set of fluxes. However, it might be possible to construct a meta-stable de Sitter vacuum out of a degenerate critical point of the flux potential by huge amount of fine-tuning. In contrast with the previous limit for which the stability requirement leads to a mild constraint on the number of non-supersymmetric vacua, the stability requirement in this limit leads to a stringent constraint. It basically prohibits all de~Sitter vacua for generic fluxes, if one does not incorporate D-term uplifting.
\par
In summary, as long as we are near to the position of supersymmetric critical points in complex structure moduli space of $Y$, it will be impossible for generic fluxes to obtain a meta-stable de Sitter vacuum by allowing the axion-dilaton to vary arbitrarily. The stability requirement may lead to entirely different constraints in different regions of moduli space.
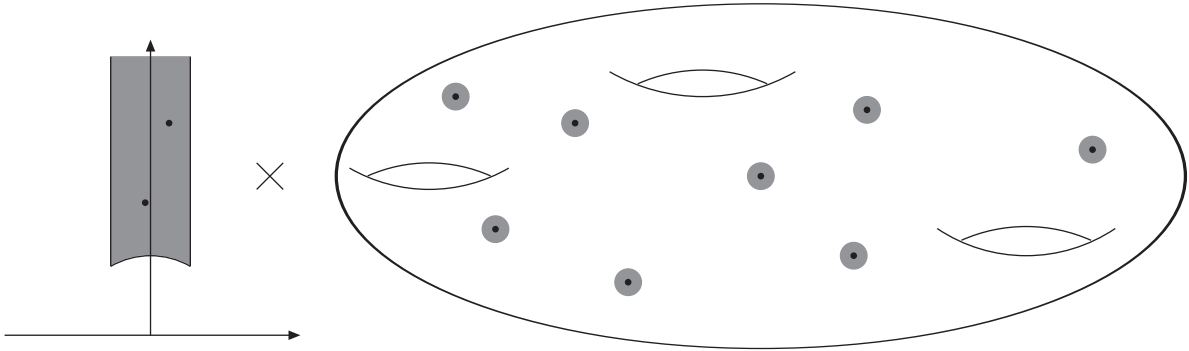
\begin{figure}[!t]
\begin{center}
\begin{picture}(300,130)(-60,0)
\CCirc(50,30){5}{Gray}{Gray}
\CCirc(80,70){5}{Gray}{Gray}
\CCirc(100,10){5}{Gray}{Gray}
\CCirc(35,80){5}{Gray}{Gray}
\CCirc(150,50){5}{Gray}{Gray}
\CCirc(190,75){5}{Gray}{Gray}
\CCirc(185,20){5}{Gray}{Gray}
\CCirc(275,60){5}{Gray}{Gray}
\CCirc(50,30){1}{Black}{Black}
\CCirc(80,70){1}{Black}{Black}
\CCirc(100,10){1}{Black}{Black}
\CCirc(35,80){1}{Black}{Black}
\CCirc(150,50){1}{Black}{Black}
\CCirc(190,75){1}{Black}{Black}
\CCirc(185,20){1}{Black}{Black}
\CCirc(275,60){1}{Black}{Black}
\Oval(150,50)(65,160)(0)
\CArc(25,105)(60,240,300)
\CArc(25,-2)(57,66,114)
\CArc(128,150)(70,240,300)
\CArc(128,30)(60,66,114)
\CArc(250,80)(60,236,304)
\CArc(250,-29)(60,66,114)
\CBox(-95,16)(-65,95){Gray}{Gray}
\CCirc(-80,-10){30}{White}{White}
\LongArrow(-135,-10)(-25,-10)
\LongArrow(-80,-10)(-80,100)
\CArc(-80,-10)(30,60,120)
\SetColor{Black}
\Line(-95,16)(-95,95)
\Line(-65,16)(-65,95)
\Line(-40,55)(-30,45)
\Line(-40,45)(-30,55)
\CCirc(-82,40){1}{Black}{Black}
\CCirc(-73,70){1}{Black}{Black}
\end{picture}
\end{center}
\vspace{0.8cm}
\caption{\textit{In this figure the black points correspond to supersymmetric critical points of the flux potential, for which} $D_{A}W=0$. \textit{The gray} $2n$-\textit{dimensional small balls on the right hand side of the picture indicate regions of complex structure moduli space of the Calabi-Yau threefold which are in the vicinity of supersymmetric critical points. On the left hand side, the gray region which corresponds to the whole moduli space} ${\mathcal{E}}$ \textit{shows that the axion-dilaton can vary freely.}}
\end{figure}
\section{Landscape of String Theory}
Motivated by string theory, there are serious attempts to formulate the space of all effective theories which can consistently couple to gravity \cite{Ooguri:2006in}. One of the main issues is to show that any effective Lagrangian cannot appear in string theory, although the landscape of string theory includes too many vacua\footnote{\ Of course, it is important to show that the number of vacua in the landscape of string theory is finite, although it might be quite big. It has been argued in \cite{Acharya:2006zw} that there are several reasons (different topologies of the extra dimensions, background fields, fluxes) that one should be worried about the finiteness of the landscape of vacua of string theory. Nevertheless, it is discussed in \cite{Acharya:2006zw} and is explicitly shown in \cite{Douglas:2006xy} for the case of intersecting brane models that the landscape of vacua is indeed finite.}. Having many explicit examples coming from string theory, several criteria have been proposed to distinguish between the acceptable effective Lagrangians and those which are not. These conjectural criteria are divided into three main categories which we briefly mention them here.
\par
First criterion gives an upper bound on the number of fields in the moduli space of the theory (or equivalently, a bound on the dimension of the moduli space) for a fixed number of dimensions of spacetime. The second criterion is an statement about the strength of the coupling constants. This conjecture states that gravity is always the weakest force or at most it is not bigger than any other force (for example, in case of extremal black holes the forces from the mass and the charge of the black hole are in balance.) \cite{Arkani-Hamed:2006dz}. The third category introduces a set of conjectures about the geometry of the moduli space. These criteria are mostly motivated by examples coming from the dualities in string theory.
\par
All these criteria are strongly sensitive to the existence of gravity and, in fact, have been designed in such away that they are wrong if gravity decouples. However, all these criteria are motivated by supersymmetric examples. It is believed that they should be true in non-supersymmetric cases as well. But in addition to these criteria, one certainly needs to impose more constraints for non-supersymmetric cases. As we saw in our case, the stability requirement imposes  additional constraints and exclude some regions of moduli space for non-supersymmetric vacua. As is clear, (\ref{eigen3}) is a nontrivial equation only for non-supersymmetric vacua. For supersymmetric ones, both sides of the equality vanish since the eigenvector vanishes. This is an explicit example in which the stability requirement imposes specific constraints for only non-supersymmetric vacua.
\section{Conclusions}
In the preceding sections, we have studied the statistics of meta-stable de Sitter vacua in specific regions of moduli space and we have addressed important features of this analysis and some other related issues. We started with orientifold limit of F-theory flux compactification on a Calabi-Yau fourfold. The four dimensional low energy effective theory is ${\mathcal{N}}=1$ supergravity. Due to the no scale structure of tree level supergravity action, fluxes can only stabilize the complex structure moduli and the axion-dilaton filed. The K\"{a}hler moduli are stabilized by incorporating KKLT type of non-perturbative effects in the superpotential. Then, the stabilization of the entire set of moduli and axion-dilaton can be divided into two steps. In the first step, one takes the axion-dilaton and complex structures of the Calabi-Yau threefold and in the next step one stabilizes the K\"{a}hler moduli by considering non-perturbative effects in the superpotential. In this manner, there are two ways to construct de Sitter vacua. Taking the axion-dilaton and complex structure moduli, one may start with a meta-stable anti de Sitter vacuum (perhaps a supersymmetric one, because solving supersymmetric equations is easier) and then stabilize the K\"{a}hler moduli. But at the end one needs to uplift the anti de Sitter vacuum to a de Sitter one by introducing a stack of anti D3-branes, which requires some amount of fine tuning in order not to destabilize the vacuum \cite{Kachru:2003aw}. Alternatively, one can start from a meta-stable de Sitter vacuum and stabilize the axion-dilaton and complex structure moduli and then fix the K\"{a}hler moduli by non-perturbative effects \cite{Saltman:2004sn}. The benefit of the later method is that it does not involve adding anti D3-branes for uplifting. Our analysis in this paper fits into the framework of the later method.
\par
We mentioned several classes of examples from \cite{Denef:2004ze} for which no meta-stable de Sitter vacua exist. If one compares the situation with ${\mathcal{N}}=2$ supergravity, it is suggestive that perhaps the Hessian of the potential has tachyonic modes. In ${\mathcal{N}}=2$ supergravity, in the absence of hypermultiplets, it has been shown that the Hessian of the potential always has tachyonic modes \cite{Fre:2002pd}, regardless of the details of the topology and geometry of the compact internal space. But in the case of flux compactification, if one takes the simplest example of Calabi-Yau for which there exists only one complex structure modulus and the axion-dilaton  field, the stability of the vacua is undetermined unless one specifies the details of the Calabi-Yau internal space. However, we showed that there exists a limit (which corresponds to certain regions in moduli space) for which the Hessian of the flux potential, regardless of the details of the topology and geometry of the Calabi-Yau internal space, is always tachyonic for de Sitter vacua for any number of complex structure moduli, as in case of ${\mathcal{N}}=2$ supergravity.
\par
We considered the limit where the supersymmetry is dominantly broken by the axion-dilaton and the contributions of F-terms associated with complex structure moduli are small ($|D_{0}W|\gg|D_{a}W|$). We explored this specific limit because of two reasons. First, the almost supersymmetric limit for which the parameters of supersymmetry breaking are very small has already been studied in \cite{Denef:2004cf}. In this limit, there is no de Sitter critical point for generic fluxes, unless one includes the D-term contribution to the potential. This, of course, has nothing to do with stability requirement. But if we allow some of the supersymmetry breaking parameters to compete with the value of the superpotential, then there is a possibility to obtain de Sitter vacua for generic fluxes without fine-tuning. Perhaps, the most naive such a pattern is to treat all complex structure moduli in the same way and keep them small but let the axion-dilaton field to compete with superpotential. Second reason arises from phenomenological interests. In four dimensional heterotic string theory, it turns out that if supersymmetry is dominantly broken by the dilaton, then all soft supersymmetry breaking parameters are universal and model independent \cite{Brignole:1993dj}. The main features of this limit are maintained in type IIB string theory as well \cite{Camara:2003ku}.
\par
In this limit (dilaton dominated), we proved that for non-degenerate critical points the holomorphic-antiholomorphic part of the mass matrix has an eigenvalue proportional to minus the sum of the value of cosmological constant and the square of the gravitino mass. This implies that the Hessian of the flux potential cannot be positive definite and always has at least one negative eigenvalue for positive critical values. For the degenerate critical points, we argued by the analogy between the mathematical structure of flux vacua and non-supersymmetric black hole attractors, it is impossible to have perturbatively stable degenerate critical points unless one does a significant amount of fine-tuning. After all, the conclusion is that there is no meta-stable de Sitter vacua in this limit. This means that the idea of treating axion-dilaton differently from complex structures does not lead to any meta-stable de Sitter vacuum without uplifting.
\par
We also studied the consequences of the results we obtained earlier on the statistics of meta-stable de Sitter vacua. In the limit of almost supersymmetric solutions, it has been shown in \cite{Denef:2004cf} that the stability requirement leads to a mild constraint on the number of non-supersymmetric vacua. However, in the limit of dominantly broken supersymmetry in the axion-dilaton direction of moduli space, the stability requirement leads to a stringent constraint. This suggests that in different regions of moduli space, the stability requirement may look completely different. There might be still other regions of moduli space in which the stability constraint imposes stringent conditions on the number of vacua.
\par
We have addressed another interesting feature of the result we obtained in (\ref{eigen3}). As is clear, the eigenvector of this eigenvalue equation is the vector whose components are supersymmetry breaking parameters. Therefore, for supersymmetric solution this equation is trivial and the constraint we found for stability is not applied for supersymmetric solutions. In other words, this is a constraint which is only applied for non-supersymmetric vacua. In the course of the current trend for finding appropriate criteria to distinguish between the landscape of meta-stable vacua and the swampland \cite{Ooguri:2006in}, this might be interesting. It is believed that the criteria stated in \cite{Ooguri:2006in} for supersymmetric vacua are valid for meta-stable non-supersymmetric solutions as well. However, it seems that, in addition, one needs to find other criteria specifically for non-supersymmetric vacua to identify their landscape.
\vspace{0.7cm}\\
\noindent{\large{\textbf{Acknowledgments}}}
\vspace{0.4cm}
\\
It is pleasure to thank F.~Denef, M.~Esole, B.~Florea, M.~Peskin, E.~Silverstein, N.~Sivanandam and, in particular,  S.~Kachru and R.~Kallosh for many stimulating and fruitful discussions. This work was partially supported by NSF grant PHY-0244728 and by US Department of Energy under contract number DE-AC02-76SF00515.
\appendix
\section{Griffiths Transversality}
\setcounter{equation}{0}
In this section, we give a proof of the Griffiths transversality based on local computations. For more technical details and related problems, one can refer to \cite{Griffith}. This proof is applicable for any compact K\"{a}hler manifold $X$ of complex dimension $k$, although we assumed that $X$ is a Calabi-Yau fourfold throughout the paper. We define a Hodge structure of weight $k$ as a finite rank lattice with the decomposition $H^{k}(X,{\mathbb{Z}})\otimes_{\mathbb{Z}}{\mathbb{C}}=\bigoplus_{p+q=k}H^{p,q}(X)$, in which the complex subspaces $H^{p,q}$ satisfy $\overline{H^{p,q}(X)}=H^{q,p}(X)$. Since the $H^{p,q}$ subspaces do not vary holomorphically inside $H^{k}$ under the variation of the Hodge structures, it is convenient to define the notion of Hodge filtration. For    a weight $k$ Hodge structure, the Hodge filtration $F^{p}$ is defined as
\begin{eqnarray}\label{filtration}
F^{p}=H^{k,0}\oplus H^{k-1,1}\oplus\cdots\oplus H^{p,k-p}\ ,
\end{eqnarray}
and it is obvious that $F^{k}\subseteq F^{k-1}\subseteq\cdots F^{1}\subseteq F^{0}=H^{k}$. In this manner, we can construct each of the $H^{p,q}$ subspaces from a Hodge filtration as $H^{p,q}=F^{p}\cap\overline{F^{q}}$. If we fix a family of $f:{\mathcal{X}}\rightarrow{\mathcal{M}}$ of compact K\"{a}hler manifolds where ${\mathcal{M}}$ is the complex structure moduli space of $X\in{\mathcal{X}}$, we can then define the holomorphic vector bundle ${\mathcal{H}}^{k}=R^{k}f_{*}{\mathbb{C}}\otimes{\mathcal{O}}_{\mathcal{M}}$, in which $R^{k}f_{*}{\mathbb{C}}$ is the  locally constant sheaf of functions on ${\mathbb{C}}$. Since each fiber of vector bundle ${\mathcal{H}}^{k}$ has a Hodge filtration, we can extend the notion of Hodge filtration on holomorphic subbundles ${\mathcal{F}}^{k}\subseteq{\mathcal{F}}^{k-1}\subseteq\cdots\subseteq{\mathcal{F}}^{0}={\mathcal{H}}^{k}$. The benefit of this definition is now that the Hodge structure varies holomorphically inside ${\mathcal{H}}^{p,k-p}\equiv{\mathcal{F}^{p}/{\mathcal{F}}^{p+1}}$.
\par
The Griffiths transversality is the relationship between the Hodge filtration and the Gauss-Manin connection defined on the above holomorphic vector bundles. First let us define the Gauss-Manin connection more precisely. The Gauss-Manin connection is a flat connection $\nabla$ on ${\mathcal{F}}^{p}$ and is locally defined as $\nabla\alpha=\sum_{i}df_{i}\otimes e_{i}$ where $\alpha=\sum_{i}f_{i}e_{i}$ is any section of ${\mathcal{F}}$ for which $\{e_{1},\cdots,e_{k}\}$ forms a frame basis ($\nabla e_{i}=0$) of sections of $R^{k}f_{*}{\mathbb{C}}$. Now,  Griffiths transversality states that the covariant derivative (with respect to the Gauss-Manin connection) of a weight $p$ Hodge filtration subspace lies inside the subspace of Hodge filtration of weight $p-1$: $\nabla{\mathcal{F}}^{p}\subseteq{\mathcal{F}}^{p-1}\otimes\Omega^{1}_{\mathcal{M}}$.
\par
In order to prove this statement, we consider a $(p,k-p)$ form on $X$ ($\rho\in\Omega^{p,k-p}(X)$). In local coordinates, we can write $\rho$ as
\begin{eqnarray}\label{rhoform}
\rho=\sum_{\tiny{\begin{array}{c}
\#I=p \\
\#J=k-p
\end{array}}}\rho_{I\bar{J}}(t_{1},\cdots,t_{k};\bar{t}_{\bar{1}},\cdots,\bar{t}_{\bar{k}})\ dt_{I}\wedge d\bar{t}_{\bar{J}}\ ,
\end{eqnarray}
where $\{(t_{i},\bar{t}_{\bar{i}});i=1,\cdots k\}$ is a chosen local coordinate system of $X$. We use capital indices to represent a multi-index set and the notation $dt_{I}$ means
$dt_{I}\equiv dt_{1}\wedge dt_{2}\wedge\cdots\wedge dt_{p}$. Now, we deform the complex structure in the direction $\frac{\partial}{\partial z_{\alpha}}$ in the tangent space of complex structure moduli space. We notice that we have chosen a local coordinate system $(z_{\alpha},\bar{z}_{\bar{\alpha}})$ for the complex structure moduli space. Here, we consider the deformation along one direction and drop the index $\alpha$. The generalization for a generic direction is easy and straightforward. If we consider a patch in neighborhood of $z$ in moduli space, then we should think of the deformation of complex structure as varying the local coordinates of $X$ as functions of $z$. Therefore, for the first order deformation we have
\begin{eqnarray}\label{changecoor}
t_{i}(z)=t_{i}+zf_{i}(t_{1},\cdots t_{k};\bar{t}_{\bar{1}},\cdots\bar{t}_{\bar{k}})+\bar{z}\bar{g}_{i}(t_{1},\cdots t_{k};\bar{t}_{\bar{1}},\cdots\bar{t}_{\bar{k}})+{\mathcal{O}}(z^{2})\ ,
\end{eqnarray}
in which $f_{i}$ and $\bar{g}_{i}$ are smooth functions defined on $X$. We assume that all second and higher order deformations are negligible. Accordingly, $\rho$ will have dependence on the coordinate $z$ of the complex structure moduli space as
\begin{eqnarray}\label{rhoformz}
\rho(z,\bar{z})=\sum_{\tiny{\begin{array}{c}
\#I=p \\
\#J=k-p
\end{array}}}\rho_{I\bar{J}}(t_{1}(z,\bar{z}),\cdots,t_{k}(z,\bar{z});\bar{t}_{\bar{1}}(z,\bar{z}),\cdots,
\bar{t}_{\bar{k}}(z,\bar{z}))\ dt_{I}(z,\bar{z})\wedge d\bar{t}_{\bar{J}}(z,\bar{z})\ .
\end{eqnarray}
The covariant derivative of $\rho$, which we want to calculate, is $\nabla_{\partial/\partial z}\rho=\frac{\partial\rho(z)}{\partial z}\Big|_{z=0}$. Therefore, we can expand $\rho$ in terms of $z$ and find the coefficient of $z$ in the expansion. We can then determine the type the covariant derivative of $\rho$ as a form. Let us to start with the components $\rho_{I\bar{J}}$
\begin{eqnarray}\label{rhoij}
\rho_{I\bar{J}}(z)=\rho_{I\bar{J}}+z\sum_{i=1}^{k}f_{i}\partial_{i}\rho_{I\bar{J}}+z\sum_{i=1}^{k}g_{i}
\bar{\partial}_{\bar{i}}\rho_{I\bar{J}}\ .
\end{eqnarray}
In the above expression, we have ignored all second and higher order terms in $z$. We have also dropped the first order  anti-holomorphic terms (which are proportional to $\bar{z}$), because after all we want to keep only terms which are first order in $z$. Expanding $dt_{I}(z,\bar{z})$ in terms of $z$ and $\bar{z}$ and keeping only linear terms, we get
\begin{eqnarray}\label{dt-expand}
dt_{I}(z)&=&dt_{I}+z\Big(\sum_{i=1}^{p}\partial_{i}f_{i}\Big)dt_{I}+z\sum_{i=1}^{p}\sum_{j=p+1}^{k}(\partial_{j}f_{i})
dt_{1}\wedge\cdots\wedge\widehat{dt_{i}}\wedge\cdots\wedge dt_{p}\wedge dt_{j}\nonumber\\
&&+z\sum_{i=1}^{p}\sum_{j=1}^{k}(\bar{\partial}_{\bar{j}}f_{i})dt_{1}\wedge\cdots\wedge\widehat{dt_{i}}\wedge\cdots\wedge dt_{p}\wedge d\bar{t}_{\bar{j}}\ ,
\end{eqnarray}
where the hatted $dt_{i}$ indicates that the $i$-th $dt$ has been eliminated. We have again dropped the anti-holomorphic terms. If we do the same thing for $d\bar{t}_{\bar{J}}(z,\bar{z})$, we similarly obtain
\begin{eqnarray}\label{dtbar-expand}
d\bar{t}_{\bar{J}}(z)&=&d\bar{t}_{\bar{J}}+z\Big(\sum_{j=1}^{k-p}\bar{\partial}_{\bar{j}}\bar{g}_{j}\Big)
d\bar{t}_{\bar{J}}+z\sum_{i=1}^{k-p}\sum_{j=k-p+1}^{k}(\bar{\partial}_{\bar{j}}\bar{g}_{i})d\bar{t}_{\bar{1}}
\wedge\cdots\wedge\widehat{d\bar{t}_{\bar{i}}}\wedge\cdots\wedge d\bar{t}_{\overline{k-p}}\wedge d\bar{t}_{\bar{j}}\nonumber\\
&&+z\sum_{i=1}^{k-p}\sum_{j=1}^{k}(\partial_{j}\bar{g}_{i})d\bar{t}_{\bar{1}}
\wedge\cdots\wedge\widehat{d\bar{t}_{\bar{i}}}\wedge\cdots\wedge d\bar{t}_{\overline{k-p}}\wedge dt_{j}\ ,
\end{eqnarray}
in which the hatted $d\bar{t}_{\bar{i}}$ has been omitted. Considering (\ref{rhoij}), (\ref{dt-expand}), and (\ref{dtbar-expand}), we can find the r.h.s. of (\ref{rhoformz}) as an expansion in terms of $z$. The coefficient of $z$ in this expansion will be the covariant derivative of $\rho$. It is then clear that all terms in $\nabla_{\partial/\partial z}\rho$ are the same type of forms as $\rho$ itself, namely $(p,k-p)$ form, except two terms. First, the term which comes from the last term in the r.h.s. of (\ref{dt-expand}) in which one $dt$ has been substituted by one $d\bar{t}$ and this makes a $(p-1,k-p+1)$ form. Second term comes from the last term in the r.h.s. of (\ref{dtbar-expand}) and this leads to a $(p+1,k-p-1)$ form. Therefore, the covariant derivative of $\rho$ includes these three types of forms $(p,k-p)\oplus(p-1,k-p+1)\oplus(p+1,k-p-1)$. Since $\rho$ is an arbitrary form which lies in  ${\mathcal{F}}^{p}$, this result implies that $\nabla_{\partial/\partial z}{\mathcal{F}}^{p}\subseteq{\mathcal{F}}^{p-1}\otimes\Omega^{1}({\mathcal{M}})$ which locally proves the statement of Griffiths transversality.

\end{document}